\documentclass[a4paper,11pt]{article}
\usepackage{aaskaiid,orcidlink}
\def\arcsec{$^{\prime\prime}$}
\def\arcdeg{$^{\circ}$}
\def\kms{~km~s$^{-1}$}
\def\uas{$\mu$as}

\title{Spatio-kinematical structure of the Galactic Nuclear Stellar Disk revealed in VLBI astrometry of circumstellar masers}
\ShortTitle{SKA-VLBI astrometry of the Galactic Nuclear Stellar Disk}

\author[1,2]{Hiroshi Imai\orcidlink{0000-0002-0880-0091}}
\ShortName{H. Imai et al.} 
\author[3]{Kohei Kurahara\orcidlink{0000-0003-2955-1239}}
\author[4,5]{Huib J. van Langevelde\orcidlink{0000-0002-0230-5946}}
\author[6,7]{Maria J. Rioja\orcidlink{0000-0003-4871-9535}}
\author[6]{Richard Dodson}

\affiliation[1]
{Amanogawa Galaxy Astronomy Research Center, Kagoshima University, 1-21-35 Korimoto, Kagoshima 890-0065, Japan}
\emailAdd{hiroimai@km.kagoshima-u.ac.jp}
\affiliation[2]{Center for General Education, Kagoshima University, 1-21-30 Korimoto, Kagoshima 890-0065, Japan}
\affiliation[3]
{Kobayashi-Maskawa Institute, Nagoya University, Furo-cho, Chikusa-ku, Nagoya 464-8602, JAPAN}
\emailAdd{kurahara.kohei.i7@f.mail.nagoya-u.ac.jp}
\affiliation[4]{Joint Institute for VLBI ERIC(JIVE), Oude Hoogeveensedijk 4, 7991PD Dwingeloo, The Netherlands}
\emailAdd{vlangeve@strw.leidenuniv.nl}
\affiliation[5]{Sterrewacht Leiden, Leiden University, Postbus 9513, 2300RA Leiden, The Netherlands}
\affiliation[6]
{International Centre for Radio Astronomy Research, University of Western Australia, 35 Stirling Hwy, Crawley, WA, Australia}
\emailAdd{maria.rioja@icrar.org}
\emailAdd{richard.dodson@uwa.edu.au}
\affiliation[7]{Observatorio Astron\'omico Nacional (IGN), Alfonso XII, 3 y 5, 28014, Madrid, Spain}

\abstract{
SKA-VLBI astrometry will enable us to measure up to thousands of three-dimensional motions of OH masers 
associated with circumstellar envelopes (CSEs) of OH/IR stars in the Nuclear Stellar Disk (NSD) and sites of 
high-mass star formation in the Central Molecular Zone (CMZ) of the Galactic Center (GC). It is expected 
that the spatio-kinematical distribution of those OH masers should indicate the existence of a ring structure in the NSD,  
which has formed as a result of outward propagation of star-formation activities in the GC. 
This is likely visualized clearly by a group of OH/IR stars, some of which should have stellar pulsation periods 
of $>$400 days and the corresponding ages of $<$500~Myr, and some sites of ongoing star formation. 
These OH/IR stars should host 1612-MHz OH masers, {some of which should become targets of huge-sample 
VLBI astrometry, in moderate accuracy,} in SKA-MID Band 2 ($\sim$1.6~GHz). 
The data of maser source proper motions will exhibit a stream motion in the stellar ring structure. 
Furthermore, the information of accurate distances (error $<$100 pc) of the maser sources are necessary to 
directly find the major-axis direction of a possible elliptical ring of stars at $\sim$8~kpc. These distances may be 
yielded through trigonometric parallaxes measurable in SKA-MID Band 5a (5--7~GHz) and/or photometric 
parallaxes derived from the pulsation period--luminosity relation of long period variable stars hosting the maser sources.}
\begin{document}
\maketitle

\section{Introduction}

High accuracy radio astrometry such as the VLBI projects conducted with VERA, VLBA, EVN, 
and LBA have revealed directly the three-dimensional, spatio-kinematical structure of the Milky Way Galaxy 
(MWG) by targeting sources of astrophysical masers from excited molecules such as CH$_3$OH, H$_2$O 
and SiO, and compact non-thermal continuum sources including 
pulsars for {\it the Galactic trigonometry} (see some reviews, e.g., \citealt{2019ApJ...885..131R,2020PASJ...72...50V}). 
However, despite such projects dedicated on VLBI astrometry, the number of target sources has been limited 
to $\sim$300 (e.g., \citealt{2024IAUS..380..111R}). This number limit has been attributed mainly to the sensitivity 
and the fields of view to observe simultaneously the targets and calibrators. 
Higher precision (up to several \uas) astrometry has been possible in a higher frequency band yielding 
higher angular resolution, such as 22 and 43~GHz covering H$_2$O and SiO masers, respectively (and CH$_3$OH 
masers at 6.7 and 12.2~GHz). In a lower frequency band, such as 1.6~GHz covering OH masers, 
the astrometric precision is improved with higher sensitivity, 
but the VLBI data are significantly affected by the Earth's ionosphere and the intrinsic structures of the target and reference 
sources (see a review of \citealt{2020A&ARv..28....6R}). 
Nevertheless, astrometry at such a low frequency still fascinates us 
because some sources such as OH masers, the main targets in this chapter, should provide key opportunities for exploring 
stellar physics and the spatio-kinematical structure of the inner part of the MWG in the era of SKA with VLBI capability. 

This chapter focuses its main scientific interests in VLBI astrometry of OH masers in the Galactic Center 
(GC), including the Nuclear Stellar Disk (NSD) whose radius is $\sim$200~pc (e.g., \citealt{2013ApJ...769L..28N}), 
and comparable to that of the Central Molecular Zone (CMZ, e.g., \citealt{2025PASJ...77L..55S}). Section 
\ref{sec:motivation} describes the background and the motivation of this science case with the SKA 
(see also \citealt{Rygl01.2026.SKA,Rioja01.2026.SKA}). 
Section \ref{sec:OH-science} 
demonstrates the scientific feasibility of the huge-sample astrometry of circumstellar OH maser sources in the NSD.  
Section \ref{sec:OH-feasibility} demonstrates the technical and operational feasibility of such astrometry in the era of 
SKA-VLBI. Here we consider SKA-VLBI as a VLBI array that is composed of the SKA providing multiple 
high-sensitivity beams from its tied-array and remote VLBI stations yielding a variety of baselines from several 
10~km with the SKA's arm stations up to several 1000~km with international stations.

\section{The NSD and CMZ hosting OH masers}
\label{sec:motivation}

\begin{figure}[t]
    \centering
         \vspace{-4mm}
	\includegraphics[width=0.90\columnwidth]{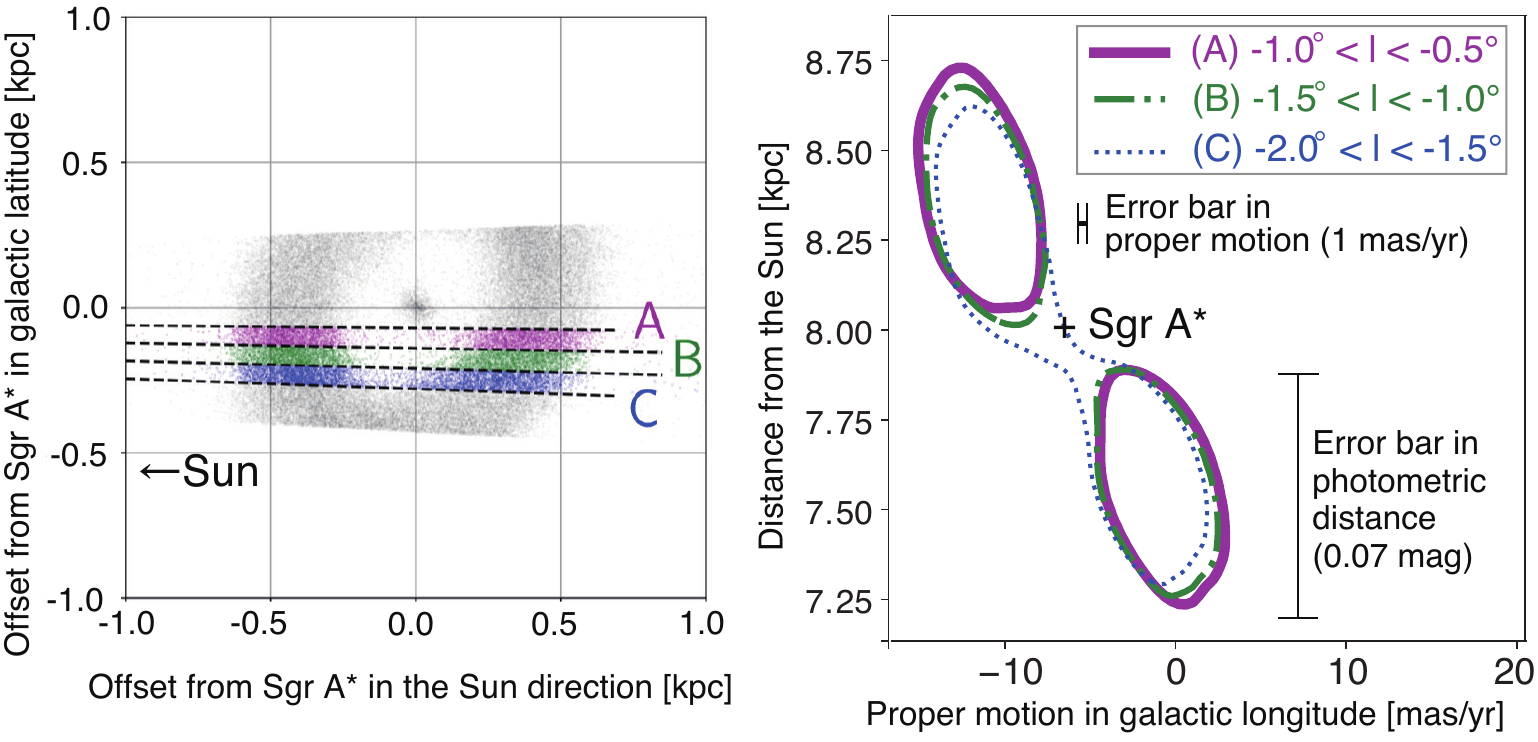}
	 \vspace{-2mm}
    \caption{{\it Left}: Face-on view of the distribution of stars within ages of 400--500~Myr in the NSD, 
    which is reproduced with the output data of the simulation by \cite{2017MNRAS.464..246B}. 
    The plotted star distribution is limited to the range of galactic longitude $-3$\arcdeg$\leq l \leq$2\arcdeg.
    Three areas, A, B, and C, are defined accordingly for the three directions 
    of galactic longitude. {\it Right}: Proper motions of stars in galactic longitude. 90\% of stars 
    in each of the areas A, B, and C are included in each enclosed area.}
    \label{fig:GC-simulation}

    \centering
        \vspace{-1mm}
	\includegraphics[width=1.0\columnwidth]{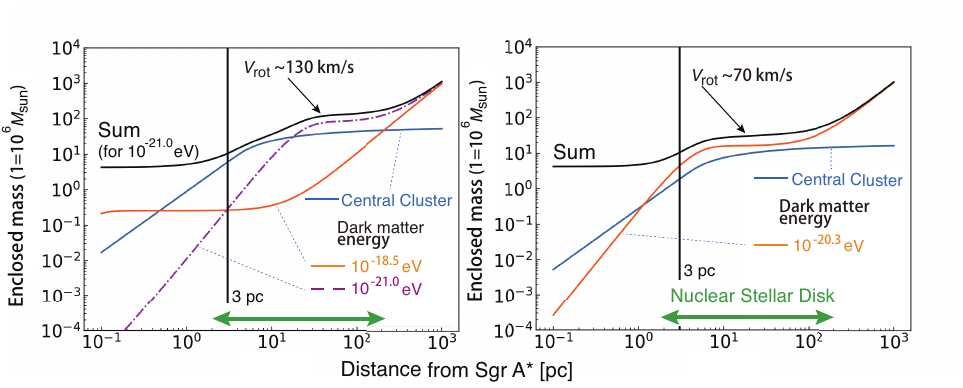}
	\vspace{-7mm}
    \caption{Mass distribution models within 1~kpc of the GC depending on energy of a dark matter particle and 
    their mass distribution in the GC presented by \citet{2022MNRAS.511.1757T}. The displayed figures are edited from 
    the original forms in the paper. The total enclosed mass, which is observationally indicated by the rotation speed, 
     is set to equal between the models within 2~pc of Sgr~A* as already observed 
    \citep{2009A&A...502...91S} and at  1~kpc from Sgr~A* regardless the energy of each 
    dark matter particle (10$^{-18.5}$, 10$^{-20.3}$, 10$^{-21.0}$ eV in the panels) dominating 
    the total mass of the NSD.       
    On the other hand, the enclosed mass and the rotation speed may change by a factor of $\sim$2 in the NSD. 
    The difference in the rotation velocity is measurable from 3D velocities of stars, 
    which should be corrected for deviations from circular motions.}
    \label{fig:ULDM}
    \vspace{-4mm}
	
\end{figure}

The MWG is confirmed to be a barred spiral galaxy, in which gas fueling into the GC has been triggered by the 
bar-shaped gravitational potential (e.g., \citealt{2025ApJ...982..185K}). Particularly, active star formation triggered by 
such bar fueling is expected to form the NSD and this will be observed directly from the orbits of stars in this region 
\citep{2020MNRAS.492.4500B}. Fig.\ \ref{fig:GC-simulation} shows the NSD visualized on basis of the simulation 
results by \citet{2017MNRAS.464..246B}, 
which exhibits a ring-shaped distribution of stars born within 400--500~Myr. Although precisely determining 
the distances from the Sun to the individual stars in the GC is challenging (e.g. \citealt{2024PASJ...76..163O}), 
the proper motions of these stars only may emerge a stellar stream along an elliptical ring structure 
as demonstrated in the right panel of Fig.\ \ref{fig:GC-simulation}. \cite{2020MNRAS.492.4500B} also suggest 
the importance of the proper motions for reducing contamination from other, kinematically hotter stars, such as 
those from the Galactic bulge, bar, and disk.

The rotation curves of the MWG, especially on the scale of the NSD (10--200~pc), also probes the distribution and 
the character of the dark matter in the GC. Fig.\ \ref{fig:ULDM} shows the MWG rotation curves and the enclosed mass 
in a few cases with different particle energies of the ultra-light dark matter (ULDM) described by \cite{2022MNRAS.511.1757T}. 

Although plenty of line-of-sight velocity information, as well as evolution and metallicity tracers, of stars and gas  
is nowadays available, the information of stellar proper motions is indispensable in order to extract the components that 
are dynamically associated with the NSD and the CMZ.  
Therefore, it is crucial to select samples of stars that are really associated with the system and trace the history 
of star formation in the system. The sample of H$_2$O and CH$_3$OH maser sources associated with the sites of 
present star formation is much limited, such as Sgr~B2, implying inactive star formation in the present CMZ. The 
measurement of the precise secular 3D motions of the interstellar maser sources is also challenging because these motions 
are significantly affected by the internal motions of the clusters of maser features associated with (high-speed) outflows 
(e.g., \citealt{2023PASJ...75..937S}). 

On the other hand, there exist hundreds of circumstellar OH (mainly at 1612~MHz) and SiO maser sources currently 
confirmed toward the GC. Deep and unbiased surveys of maser sources have been conducted toward small areas around 
(within 20\arcsec--0.4\arcdeg or 0.8--60~pc of) Sgr~A* (e.g., \citealt{1998A&AS..128...35S,2002A&A...391..967S}). 
Compared with $\geq$6 000 stars within 1~pc of  Sgr~A* with measured proper motions \citep{2009A&A...502...91S}, 
the number of stars whose proper motions are determined in SiO maser astrometry is quite limited 
($\leq$35, \citealt{2025PASJ..tmp...59T}). In wider sky areas corresponding to the NSD, SiO 
maser surveys have been conducted, yielding hundreds of detections towards targeted infrared stars (e.g., 
\citealt{2004PASJ...56..261D,2006PASJ...58..529F}; Tsuboi et al.\ in private communication). 
Note that the Bulge Asymmetries and Dynamic Evolution (BAaDE) 
survey \citep{2024IAUS..380..292S} covered a larger area and targeted $\sim$28 000 stars, 
but yielded a limited number of maser detections in the NSD. 
We note the pioneering work by \cite{1992A&A...259..118L}, who demonstrated the NSD rotation using 
a sample of $\sim$130 OH/IR stars.

\begin{figure}[b]
    \centering
         \vspace{-3mm}
	\includegraphics[width=0.90\columnwidth]{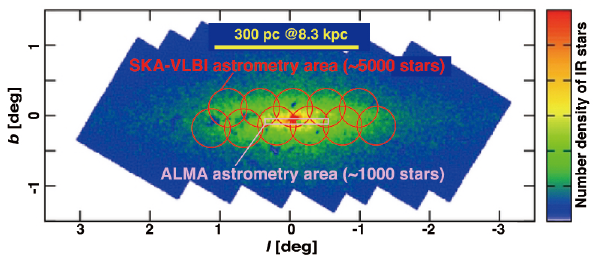}
    \vspace{-3mm}
    \caption{Areas of the SKA-VLBI astrometry toward the NSD. A red circle indicates the field of view or the single-dish beam of 
    SKA-MID Band 2 ($\sim$0.5\arcdeg) in the VLBI astrometry, which covers a region of a moderate stellar density shown in 
    the background colored map of the number density of near-infrared stars \citep{2013ApJ...769L..28N}. For comparison, 
    a magenta box indicates the area of the ALMA astrometry that aims to measure $\sim$1000 proper motions of SiO maser 
    sources within a field of currently realistic, general observation programs.}
    \label{fig:GC-astrometry-field}
\end{figure}

Here it should be highlighted that these masers are associated with long period variable stars (LPVs), many of which have 
been identified through infrared photometry (e.g., \citealt{2001MNRAS.321...77G,2009MNRAS.399.1709M}). 
The ``multi-epoch infrared VISTA Variables in Via Lactea (VVV)'' survey \citep{2022MNRAS.517..257S} has identified 
1782 LPVs in an area of $3\times3$~deg$^2$ of the GC. The ages of these LPVs may be well 
determined in their period--age relation (e.g., \citealt{2022A&A...658L...1T}), in which stars with $P\gtrsim 400$~days may 
have $t_{\rm age}\lesssim$500~Myr. Such LPVs often host SiO masers (e.g., \citealt{2004PASJ...56..261D}). Note that OH 
masers are also hosted by such LPVs although their detection rate is lower than that of SiO masers 
\citep{2004PASJ...56..261D}.
OH maser source surveys recently conducted will be deeper and cover wider sky areas than the previous ones 
mentioned above, such as the GASKAP-OH 
(Galactic ASKAP spectral line survey for hydroxyl, \citealt{2024IAUS..380..486D}), so as to detect thousands of OH masers. 

Fig.\ \ref{fig:GC-astrometry-field} shows the survey area of circumstellar OH maser sources described in this chapter. 
Extrapolating the result of the recent SiO maser astrometry with ALMA in a small area around Sgr~A* \citep{2025PASJ..tmp...59T}, 
one can estimate to measure $\sim$1000 proper motions of SiO masers in a larger area shown in Fig.\ 
\ref{fig:GC-astrometry-field}. If the astrometry of OH masers is conducted in an area $\sim$20 times 
as large as that for SiO masers, proper motions of up to $\sim$5000 OH masers may be measurable. 
Taking into account the SKA-VLBI beam ($\sim$3\arcsec) much smaller than the single beam of the SKA-MID antennas 
or the field of view in the source finding surveys ($\sim$0.5\arcdeg), the targets of the VLBI astrometry should be 
selected effectively after the surveys. 

\section{Huge sample VLBI astrometry of circumstellar OH masers in the NSD}
\label{sec:OH-science}

The above consideration on the measurement of OH maser proper motions supposes OH masers brighter than 50~mJy at a 
detection over 7-$\sigma$ (here $\sigma$ is an rms noise level) in VLBI baselines ($\gtrsim$5000~km) as a target of 
the SKA-VLBI astrometry. This level of sensitivity 
is comparable to or a little higher than those toward the GC in the previous VLA/ATCA survey \citep{1998A&AS..128...35S} 
and the ongoing GASKAP-OH survey \citep{2024IAUS..380..486D} when assuming a velocity width of $\sim$1~\kms. 
Here we discuss the sensitivity of SKA-VLBI that should be feasible for the VLBI astrometry.
Based on the output of {\it the SKA-MID Sensitivity Calculator}, we assume an SEFD$\sim$4~Jy at 1.6~GHz (Band 2) in 
the AA4 phase for the phased-up SKA-VLBI beams, around the middle between those in the AA* and AA4 phases. 
Note that the difference between the SEFDs assumed here and in the AA4 phase is attributed to the sub-arrays 
simultaneously observing calibrators together with the targeted OH masers in the Multi-View (MV) technique
\citep{2020A&ARv..28....6R} mentioned later. As possible remote VLBI stations, 
we here suppose the 26-m telescope in Hartebeesthoek, South Africa (SEFD$=$430~Jy), 65-m in Sardinia, Italy 
($\sim$100~Jy at an elevation of $\sim$20\arcdeg), 32-m in Noto, Italy ($\sim$900~Jy at $EL\sim$25\arcdeg), 
40-m in Chiang Mai, Thailand ($\sim$100~Jy), 26-m in Hobart ($\sim$500~Jy) and ATCA ($\sim$80~Jy) in Australia 
although other EVN antennas also may be available.   
Assuming a bandwidth of 5~kHz ($\simeq$0.9\kms) for OH maser detection and 2-bit quantization in signal recording, 
we calculated the 1-$\sigma$ noise level to be $\sim$10~mJy in integration of 1.5 hours and $\sim$7~mJy in 3 hours. 
The integration for $\sim$3 hours may be realistic in the SKA-VLBI toward the GC because of the sub-array capability of 
the SKA-VLBI and the common sky toward the GC with the remote stations. In this case, with a signal-to-noise ratio of  
$\sim$7-$\sigma$, thermal limits will give $\sim$1~mas accuracy astrometry.  

In order to cover the fields of astrometry (11 single 15-m antenna beams) as shown in Fig.\ \ref{fig:GC-astrometry-field}, one 
will need 11 VLBI observation sessions per year, each for a single 15-m antenna beam tracked for $\sim$4 hours 
(including overhead of $\sim$1~hour). 
Planning a project of 5 years for visiting each field in 5 times, one will request 220 hours in total. 
Thus, proper motions of OH masers may be determined in accuracy of $\sim$0.3~mas~yr$^{-1}$ 
($\sim$10\kms\ at the GC), sufficient for finding the spatio-kinematics of the NSD predicted in Fig. \ref{fig:GC-simulation} 
and \ref{fig:ULDM}. However, note that the SKA-VLBI beam is tiny ($\sim$3\arcsec) as already mentioned; therefore, 
one will request simultaneously multiple SKA tied-array beams towards the OH masers. Eventually, the total number 
of OH masers targeted for VLBI astrometry will be severely limited and proportional to the number of the SKA tied-array 
beams simultaneously available in narrow widths of base-band channels. 

As long as a single field of view is tracked continuously over 3 hours, this project may have flexibility for joining commensally 
with other surveys toward the GC. Even in running alone the astrometry project proposed here, using all the available SKA 
antenna is essential for mapping the whole regions of circumstellar OH masers in high dynamic range, covering the emission 
extended up to $\sim$1\arcsec ($\sim$8000~au at the GC). In fact, the angular size of the emission region can be compared 
with a linear size of the emission along the line of sight to derive geometrically the distance to the OH maser source 
(so-called phase-lag method, \citealt{1990A&A...239..193V}). The luminosity distances to the targeted LPVs hosting the 
OH masers may also be determined by combination of the data of infrared-photometry such as VVV 
\citep{2022MNRAS.517..257S} and trigonometric parallax distances to the LPVs in the solar neighborhood for providing 
their precise $P$--$L$ relation in the MWG \citep{2024evn..conf..137N}. 

\section{Feasibility of high precision VLBI astrometry of OH masers}
\label{sec:OH-feasibility}

VLBI trigonometry has been conducted for circumstellar OH masers nearby the Sun (e.g., \citealt
{2000A&A...357..945V, 2017AJ....153..119O}). The results of such OH maser astrometry in early time 
were affected by the Earth's ionosphere and the accuracy has been limited to $\sim$1~mas. \cite{2017AJ....153..105R} 
proved that the MV technique is really feasible to improve the precision of OH maser astrometry. 
If this technique is fully applied to the SKA-VLBI project proposed here, some of the OH masers 
($\sim$10 even in the small area covered by \citealt{1998A&AS..128...35S}) brighter than 800~mJy may be the target for 
trigonometry in 10-$\mu$as accuracy \citep{2020A&ARv..28....6R,Rioja01.2026.SKA}. 
Even the maser proper motions in 1-mas~yr$^{-1}$ accuracy may better determine the rotation curve of 
the MWG on the NSD scale (Fig.\ \ref{fig:ULDM}).

\begin{table}[h]
	\centering
	\caption{Reference source candidates around the astrometric survey field.}
	\label{tab:reference-sources}
	\begin{tabular}{llccr} 
		\hline
		ID & Name & R.A. (J2000.0) & Decl. (J2000.0) & Detected band \\
		\hline
		1& J1745$-$2820 & 17:45:52.494523 & $-$28:20:26.28551 &  XK \vspace{-1mm} \\
		2 & J1752$-$2956 &  17:52:33.108069 & $-$29:56:44.91563 & SX \vspace{-1mm}\\
		3 & J1752$-$3001 & 17:52:30.950082 & $-$30:01:06.68422 & SX \vspace{-1mm}\\
		4 & J1743$-$3058 & 17:43:17.886798 & $-$30:58:18.65557 & SCX \vspace{-1mm}\\
		5 & J1736$-$2737 & 17:36:10.177651 &  $-$27:37:20.22288 & CX \vspace{-1mm}\\
		6 & J1754$-$3031 & 17:54:56.768512 & $-$30:31:44.12038 & SCX \vspace{-1mm}\\	
		7 & J1756$-$2807 & 17:56:49.656940 & $-$28:07:37.67697 & S \vspace{-1mm}\\
		8 & J1758$-$3029 & 17:58:22.651936 & $-$30:29:15.79542 & CX \vspace{-1mm}\\
		9 & J1731$-$3003 & 17:31:46.851363 & $-$30:03:08.95426 & CX \vspace{-1mm}\\	
		10 & J1751$-$2524 & 17:51:51.262535 & $-$25:24:00.06409 & SCX \vspace{-1mm}\\
		\hline
	\end{tabular}
\end{table}

\begin{figure}[h]
\begin{minipage}{6.5cm}
	\includegraphics[width=1.0\columnwidth]{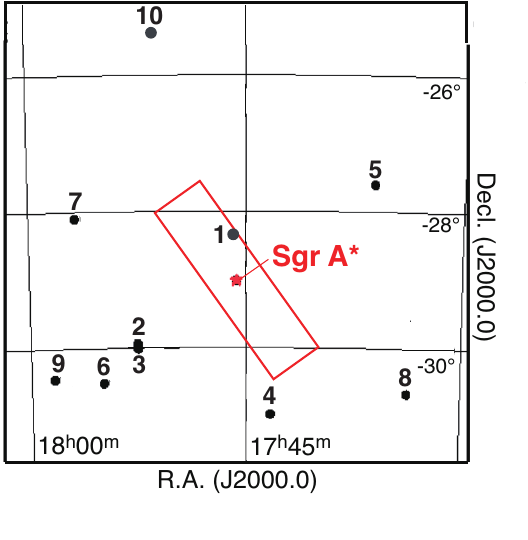}
	\vspace{-13mm}
\end{minipage}
\begin{minipage}{8.3cm}
    \vspace{-6mm}
    \caption{Distribution of the possible reference sources around the field of the planned SKA-VLBI astrometry toward 
    the NSD. The IDs of the reference sources are listed in Table \ref{tab:reference-sources}. 
    The area of the astrometry is roughly drawn in a red box.}
    \label{fig:GC-references}
\end{minipage}
\end{figure}

Table~\ref{tab:reference-sources} gives the list of the possible reference sources that are catalogued in the 
{\it VLBI Calibrator Search}\footnote{http://astrogeo.org/calib/search.html} as of 2026 March 
and have unresolved flux densities (in a baseline length of 20~M$\lambda$ or 2500~km in S-band) 
higher than 20~mJy in S (2.3~GHz)-, C (5.0~GHz)-, or X (8.4~GHz)-band, suitable for L (1.6~GHz)-band VLBI astrometry. 
Fig.~\ref{fig:GC-references} shows the distribution of these sources, 
which well bracket the target area of the astrometry (Fig. \ref{fig:GC-astrometry-field}) realizing the MV technique. 

We find 4 compact S-band continuum sources catalogued within 2\arcdeg\hspace{-2pt}.5 of Sgr~A*, 
which have some unresolved flux densities but do not well bracket even Sgr~A* alone. 
In order to find such compact sources that well bracket the astrometry field, 
we need a separation angle of up to 5\arcdeg\hspace{-2pt}.5 from Sgr~A*. 
In this case, we find 10 compact sources detected in either S- or X-band. 
Thus the MV technique is essential to cover a relatively large separation angle (up to 7\arcdeg) between 
the calibrators. 

Here note that circumstellar OH masers have their large intrinsic sizes mentioned in Sect. \ref{sec:OH-science} 
and those towards the GC are also affected by strong interstellar scattering (\citealt{1992ApJ...396..686V} and 
the references therein). 
Although some of the individual velocity components of OH masers located closely to the Galactic mid-plane are 
spatially unresolved in VLBI (e.g. \citealt{2013ApJ...773..182I}), we need VLBI baselines including the SKA tied-array  
with moderate lengths ($\sim$1000~km).  Such SKA-VLBI baselines will be available with only Hartebeesthoek 
in the early phase of the SKA. Therefore, our science case described here focuses the main goal 
on measurement of a limited fraction of proper motions of the OH masers detectable with the SKA. 

Higher precision of astrometry is expected in higher frequency bands such as SKA-MID Band 5a (5--7~GHz), 
Band 6 ($\sim$22 GHz), and 7 ($\sim$43 GHz) for circumstellar OH ($\sim$6.0~GHz), H$_2$O 
($\sim$22~GHz) and SiO ($\sim$43~GHz) masers. SKA-MID Band 6 and 7 will be available 
in the future SKA Observatory Development Program. If trigonometric parallax distances of the maser sources 
are determined for tens of stars hosting those masers in an accuracy better than $\sim$100~pc, it may be 
possible to directly find the major-axis direction of a possible elliptical ring of stars at $\sim$8~kpc 
(see Fig.\ref{fig:GC-simulation}). 

However, the fields of view are smaller and the coherent times are shorter than those in Band 2, 
which should be taken into account for operation of remote stations that may need fast antenna nodding for source switching. 
Therefore, the target sources for such astrometry should be carefully selected in the bright OH maser sources in Band 2, 
to which the demonstration of the trigonometry may be still possible with the data set of the proper motion measurement. 

Finally, the authors emphasize that the AA* phase should be a key to technically demonstrate any issue in the 
VLBI astrometry proposed here, starting with the beam that includes the calibrator candidate closest to Sgr~A*, 
J1745$-$2820, and the survey area of OH masers (see Fig.~\ref{fig:GC-references}). Involvement of the future African 
VLBI Network, including the Ghana 32-m telescope that is recently operational in scientific observations  
(\citealt{2026JATIS...12..017001P}), also will improve the astrometric accuracy.
\\

{\it Acknowledgments:} We thank K.~Hattori for providing the plots of Fig.~\ref{fig:GC-simulation}, with our edition 
dedicated for this chapter. This work was supported by the JSPS Bilateral Collaboration Program (ID:120239936) 
and the NINS International Collaborative Research Program (P.I. M.~Honma). 

\bibliographystyle{abbrvnat}
\bibliography{chapter_HImai}

\end{document}